# A State-of-Knowledge Review on the Endurance Time Method

Homayoon E. Estekanchi[1], Mohammadreza Mashayekhi[1], Hassan Vafai[2], Goodarz Ahmadi[3], Sayyed Ali Mirfarhadi[1] and Mojtaba Harati[4]

## Abstract

Endurance time method is a time history dynamic analysis in which structures are subjected to predesigned intensifying excitations. This method provides a tool for response prediction that correlates structural responses to the intensity of earthquakes with a considerably less computational demand as compared to conventional time history analysis. The endurance time method is being used in different areas of earthquake engineering such as performance-based assessment and design, life-cycle cost-based design, value-based design, seismic safety, seismic assessment, and multicomponent seismic analysis. Successful implementation of the endurance time method relies heavily on the quality of endurance time excitations. In this paper, a review of the endurance time method from conceptual development to its practical applications is provided. Different types of endurance time excitations are described. Features related to the existing endurance time excitations are also presented. Particular attention is given to different applications of the endurance time method in the field of earthquake engineering.

**Keyword**: Endurance time method, time history analysis, seismic response assessment, performance-based design, value-based seismic design.

## Nomenclature

| | |
|---|---|
| $a_g(\tau)$ | acceleration time history of an Endurance Time excitation |
| ET | endurance time |
| ETEF | endurance time excitation function |
| g (t) | a function pertinent to the intensifying profile of an ETEF |
| $S_a(T,t)$ | acceleration spectra produced by ETEFs at time t and period T |
| $S_u(T,t)$ | displacement spectra produced by ETEFs at time t and period T |
| $S_{aT}(T)$ | target acceleration spectrum |
| $S_{uT}(T)$ | target displacement spectrum |
| $T_{max}$ | maximum period of vibration in the simulation procedure |
| $t_{max}$ | motion duration of an ETEF |
| $t_{target}$ | Target time of an ETEF |
| $T_{min}$ | minimum period of vibration in the simulation procedure |
| $\ddot{x}(\tau)$ | acceleration response of an SDOF |
| $x(\tau)$ | displacement response of an SDOF |

[1] Department of Civil Engineering, Sharif University of Technology, Tehran, Iran
[2] Department of Civil Engineering and Engineering Mechanics, The University of Arizona, Tucson, AZ, USA
[3] Department of Mechanical and Aeronautical Engineering, Clarkson University, Potsdam, NY, USA
[4] Department of Civil Engineering, University of Science and Culture, Rasht, Iran

# 1. Introduction

For seismic analysis of new or existing structures, seismic codes such as ASCE-07 (2010), rehabilitation provisions (e.g. FEMA-356 (2000) and ASCE/SEI 41-17 (2017)) typically recommend several frameworks, including Linear Static Procedure (LSP), Linear Dynamic Procedure (LDP), Nonlinear Static Procedure (NSP), and Nonlinear Dynamic Procedure (NDP). Each of these procedures has its own merits and advantages. For example, the LSP and NSP are quite fast compared to the other frameworks and can be readily used by practicing design engineers. However, they are not capable of satisfactorily incorporating the dynamic characteristics of ground motions, and therefore, do not account for the potential complicated seismic effects in the structural responses. To this end, the LDP can include the effects of earthquakes in terms of their dynamic characteristics, but it is incapable of considering significant nonlinearities in the structure (Chopra 1995; Bozorgnia and Bertero 2004).

On the other hand, the NDP is capable of considering nonlinearities that arise both from materials and structural elements. While NDP is not as fast as linear frameworks and is a rather time-consuming process, it incorporates the dynamic nature of the earthquakes, as well as, the nonlinear structural behavior; thus, it is the most reliable framework in the field of earthquake and structural engineering. A NDP can be used for design of structures with complex behavior or for structural retrofitting. Examples of the cases that justify the application of the NDP procedure include base-isolated buildings and structures equipped with vibration control devices.

Other frameworks, such as Cloud Analysis (CLA) and Incremental Dynamic Analysis (IDA), have also been introduced for examining the structural behavior through several seismic levels up to the point recognized as the collapse point of the structural system. The IDA framework (Vamvatsikos and Cornell 2002), known as the most comprehensive and reliable dynamic analysis, is one of those incremental dynamic procedures through which structures are subjected to a multitude of NDPs. In order to reduce the uncertainties associated with the results of this procedure, several appropriate earthquake ground motion records are first carefully selected. Then, each ground motion record is scaled from a relatively lower intensity measure (IM) up to a level that may cause a complete collapse or dynamic instability of the considered structure. The IM is typically selected to be the spectral acceleration of the records at a range or specific natural vibration period. For each one of those considered ground motions, and at each level of intensity measure (or at each scaled level of ground motion), the NDP is used for the complete IDA analysis. Typically, several NDPs are performed for a selected earthquake since multiple seismic scaling levels are used in this approach. Therefore, the IDA procedure is very time-consuming for most practical engineering design routines.

Endurance time method is a rather fast incremental-based dynamic time history analysis in which structures are subjected to intensifying base acceleration loading. This method offers structural response predictions in terms of the relationship between engineering demand parameters (EDPs) and intensity measures (IMs). Engineering demand parameter describes structural responses while intensity measures are related to the intensity of earthquakes at different seismic levels. In the ET method, a single time history analysis provides performance status of the structure for a continuous range of IMs while computational outputs of the conventional time history analysis are only valid for a particular IM level. In fact, structural responses at several levels of intensity measures, as it is provided by an IDA, are covered by the ET method using a minimal number of analyses (typically about three).

In the present study, a review of recent advances in the development of the endurance time method is presented. The basic concept of the ET method is described, and particular attention is given to the generation techniques of endurance time excitations. It is emphasized that the reliability of the ET method results depends on the quality and properties of the endurance time excitations used. The important features of the available endurance time excitations are presented and discussed. Finally, applications of the ET method in different areas of earthquake engineering are described.

## 2. Endurance Time Method Concept

The basic concept of the ET method is inspired by the so-called stress test in medical science. In this exercise test, patients run on a treadmill and their health indicators such as blood pressure and heart signals are monitored, while the speed and slope of treadmill gradually increase during the test. Doctors judge the patient's cardiovascular system status based on the maximum endured intensity based on recorded indicators. The exercise test concept is relatively simple considering the complexity of the human body system. Considering the fact that even the most complicated structures do not have a fraction of the complexity of human bodies, a similar concept should also be applicable in their case. This motivated the concept of "Endurance Time" framework for evaluating the seismic performance of structures (Estekanchi et al. 2004). In this approach, the structures are subjected to continuously intensifying dynamic excitations and their seismic performance is assessed based on their response at applied levels of equivalent seismic intensity.

The concept of endurance time method can be explained by a hypothetical shake table test. In this example, the objective is to determine the relative performance of three structures under earthquake excitation. These structures are placed on a shaking table as shown in Figure 1 and are subjected to an intensifying dynamic excitation.

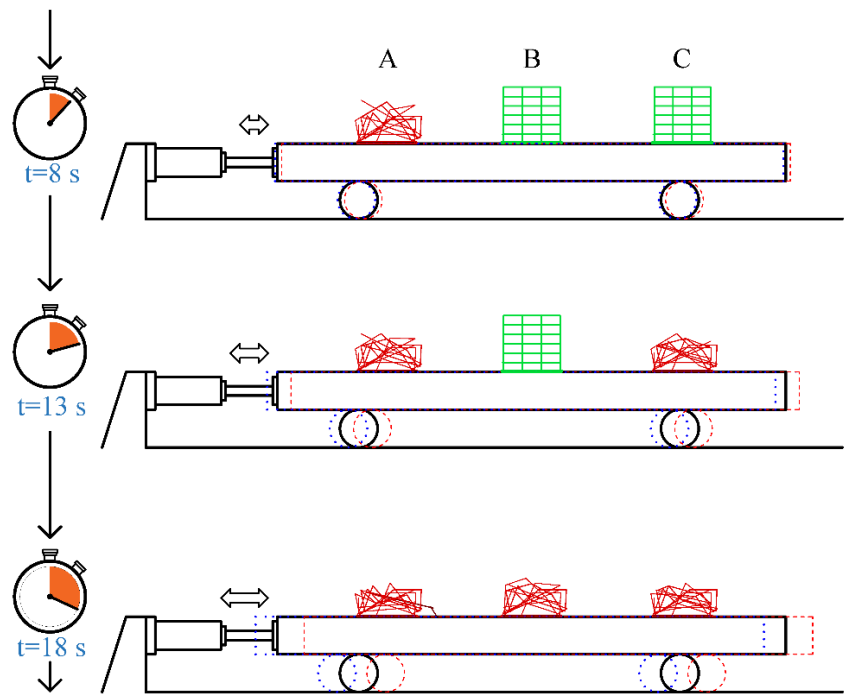

Figure 1. The concept of the Endurance Time method for determining the seismic endurance

A sample intensifying excitation is displayed in Figure 2. The increasing trend of ET acceleration function gives a new meaning to the time in the ET method; time in the ET method reflects intensity measures (IMs) of earthquake motions. At the beginning of endurance time excitations, the intensity of motions is low and hence endurance time excitations at initial time intervals are representative of low intensity earthquake ground motions. At the middle time interval of endurance time excitations, the intensity of motions is moderate and therefore excitations are representative of moderate earthquakes. At the end of endurance time excitations, the intensity is high and ET excitations are representative of severe earthquakes. In other words, time is an intensity indicator in the endurance time method.

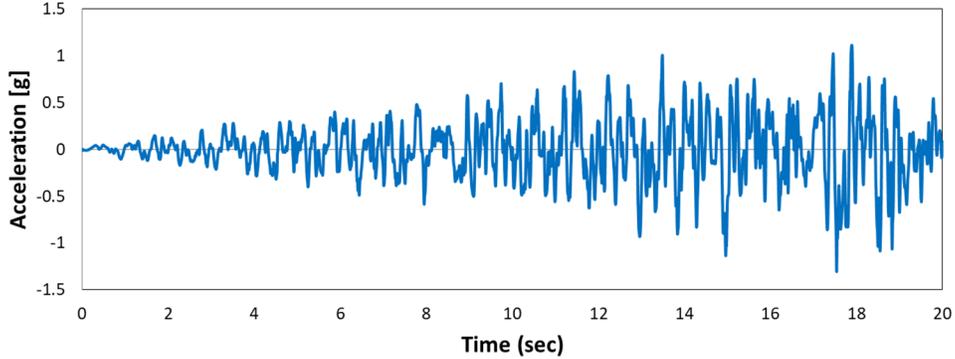

Figure 2. A sample of intensifying ET excitation, the ETA20f01 (Estekanchi 2019)

As the amplitude of the ET excitation increases in time, the structures are expected to move gradually from an elastic state to a nonlinear (plastic) state and finally collapse. Damage indicators such as maximum inter-story drift ratio of these structures are monitored during the test and reported through an endurance time curve. A sample endurance time analysis curve relating maximum absolute response parameter and the time is depicted in Figure 3. In this figure, the vertical axis of these ET curves are a measurable structural response parameter, e.g. here the $V/V_{Design}$ denotes to the relative limit state of a damage indicator for collapse prevention (CP) status of the considered structures. The values related to the vertical axis of ET curves are computed as follow:

$$\Omega(f(t)) = Max(|f(\tau)|) \quad 0 \leq \tau \leq t \qquad (1)$$

In the above equation, $\Omega$ is the maximum absolute response in the time span [0, $t$] and $f$ is the response history as a function of time. Any response or damage indicator of interest such as maximum drift, base shear and plastic rotation can be considered as response-related parameters. In view of Equation (1), if maximum inter-story drift is taken as a response parameter, $\Omega$ is the maximum drift ratio that the structure experienced during a time interval from 0 to time $t$. So maximum seismic demand of structures can be found for different intensity levels, which is a function of intensity measure itself since time and intensity measures are correlated to each other in ET method. Hence, structural responses at different individual seismic intensity levels can be determined in a single time history analysis within the endurance time framework, considerably reducing the computational demands encountered in such analyses.

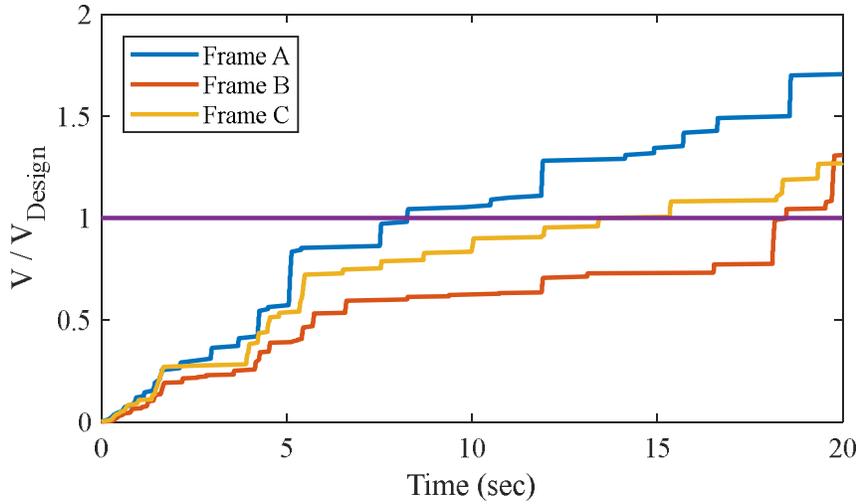
Figure 3. Increasing response plot (or ET curve) of three structures, A, B and C

As can be seen from Figures 1 and 3, structure A failed first and hence has shown the minimum seismic resilience among the other structural models being considered in this hypothetical shake table experiment. On the other hand, structure B failed after structure A and C, therefore, structure B has shown the highest seismic endurance. In addition to ranking these structures, the damage capacity of the buildings can also be quantified. In this case, Figure 3 provides a damage indicator (the relative CP limit state for each structural system or the $V/V_{Design}$) versus analysis time for the above-mentioned models, which is obtained through the hypothesized shake table test within the endurance time framework. It displays that structure A, B and C, respectively, collapse at 8sec, 18sec, and 13sec. In fact, the damage capacity of a specific structure can be defined by the maximum time that the structure can endure the input ET excitation. So, if these three structures are designed for the same seismic design target, structure B shows that it has the highest design ratio in term of seismic lateral strength. Overall, the seismic performance of structures can be quantified by their endurance time (or its equivalent seismic intensity level).

The deductions made about the ET curves in Figure 3 are according to a rather simple but direct observation that was made about the dynamic behavior of candidate structures on an imaginary shake table. The ET approach provides information on the seismic performance of a structure and compares it with those of other structures. In addition, if the input excitations are to be calibrated such that they fulfill the code design requirements, a minimum acceptable target endurance time, as described in the next section, can be set for a given seismic load. Therefore, seismic performance of a specific structure to a specific seismic excitation can be simply evaluated by comparing the demand endurance time versus the target time that is computed at the intensity level of interest.

## 3. Generation of Endurance Time Excitations

Successful implementation of the ET method depends on the quality of the endurance time excitation functions (ETEFs). Simulating more accurate endurance time excitations is a fundamental step in improving the effectiveness of the ET method. Endurance time excitations must be generated so that they can be used to predict real ground motion effects. Compatibility with real ground motions and intensification are two main features of endurance time excitations that must be considered in the simulation. The concept of response spectra can be used to simulate ET acceleration functions and provides a good starting point (Estekanchi et al 2004). In fact, response spectra of the original endurance time excitations were set to increase with time while they remained consistent with the design spectra or the response spectra of real ground motions. One simple approach is to consider the product of a target spectrum and an intensifying function; therefore, the response spectra of the endurance time excitation can be computed by multiplication of the target spectrum and the intensifying function that defines the scale factor as a function of time. In this way, the shape of acceleration spectra of endurance time excitations will be the same at all times and only the amplitude of acceleration response spectra changes. In this approach, response spectrum of intensifying ET accelerations will be only a function of time and the shape of a fixed target spectrum, where the target spectrum may be the average acceleration spectra of a ground motion suite—a far-field set (Riahi et al. 2010) or a near-field one (Ghahramanpoor et al. 2015)—or a design code acceleration spectrum, e.g. the design spectrum of ASCE07 (2010). Acceleration and displacement spectra of endurance time excitations can then be expressed as:

$$S_{aT}(T,t) = g(t) \times S_{aT}(T) \quad (2)$$

$$S_{uT}(T,t) = g(t) \times S_{uT}(T) \quad (3)$$

where $S_{aT}(T,t)$ and $S_{uT}(T,t)$ are target acceleration and displacement response spectra of endurance time excitations at time $t$, and structural natural period at first vibration mode, $T$. $S_{aT}(T)$ and $S_{uT}(T)$ are the corresponding target acceleration and displacement spectra. In this case, $g(t)$ is the intensifying function which is an ascending function of time (e.g., a linear, exponential or other ascending function). Whereas there is no limitation for intensifying function except for continuously increasing condition, in currently available endurance time excitations, linear and exponential intensifying functions have been adopted. In the generation of early endurance time excitations, the linear profile was employed. The overarching advantage of this form was its simplicity. When using linear intensification, the acceleration spectrum at 20sec is simply twice the corresponding acceleration spectrum at 10sec. Similarly, an acceleration spectrum at 5sec is half of the corresponding acceleration spectrum at 10sec. Other intensification profiles can also be applied. For example, by introducing exponential profile a cumulative absolute velocity (CAV) consistency for production of ET excitations can be achieved (Mashayekhi et al. 2018a). The linear and exponential intensification profile forms are given as:

$$g(t) = t/t_{target} \quad (4)$$

$$g(t) = b \tanh(\gamma t) e^{\alpha t} \quad (5)$$

where $t_{target}$ is the time at which endurance time excitations produces the target acceleration spectrum. For brevity, $t_{target}$ is called *target time*. In the linear form of the intensifying function given by equation 4 the target time is directly included in the formula. In the corresponding exponential form of the intensifying function given by Equation (5), the target is adjusted through assigning values to constant parameters b, $\gamma$ and $\alpha$. It should be mentioned that a target time of 10sec was typically used at the early stage of the ET method development, which was based on engineering judgment. In the production of 40sec CAV-consistent endurance time excitations, the target time to reach a scale factor of unity, is adjusted to be equal to 20sec.

A sample endurance time excitation is shown in Figure 4 (a). This excitation is selected from "ETA20in" series of endurance time excitations for which a linear intensification function is used (Estekanchi 2019). Different series of endurance time excitations and their characteristics will be discussed in Section 4. From Figure 4 (b) in which acceleration spectra of ETA20inx01 are depicted, it can be seen that the acceleration spectrum at the target time of t=20sec is approximately twice the acceleration spectrum at t=10sec. In addition, the acceleration spectrum at t=15sec is 1.5 times the acceleration spectrum at t=10sec. Similarly, the acceleration spectrum at t=5sec is half of the acceleration spectrum at t=10sec. As can be seen from Figure 4, a single ET excitation record is simulated in a way that it produces different predefined target spectra at different relevant target times.

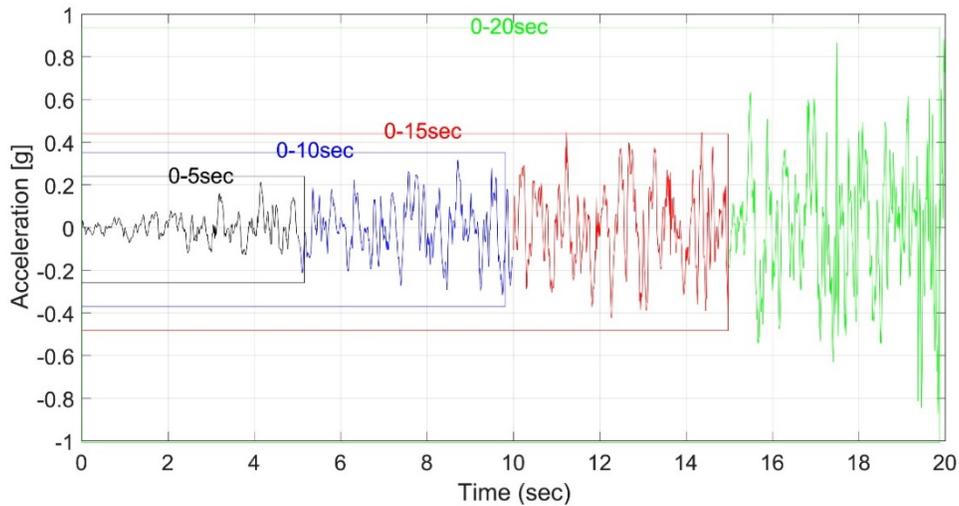

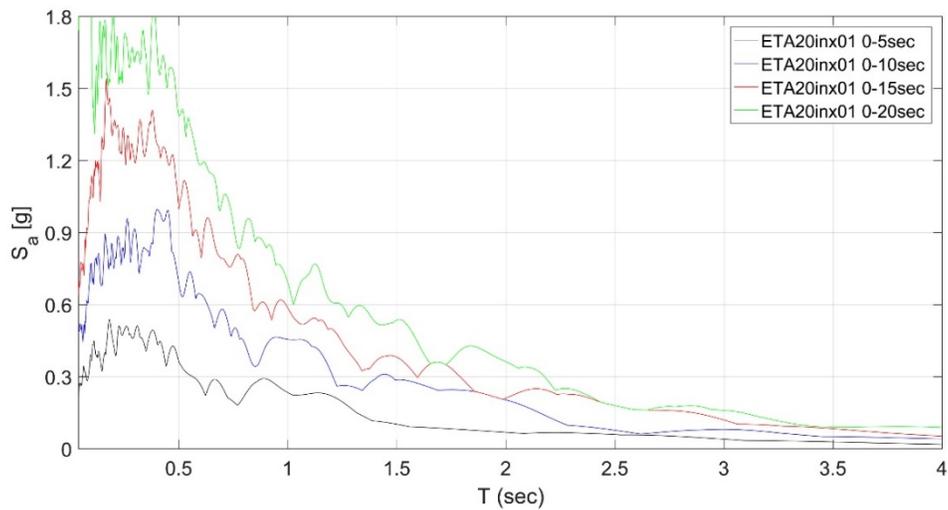

Figure 4. (a) ETA20inx01 acceleration time history, (b) acceleration spectra of ETA20inx01 at times 5sec, 10sec, 15sec, and 20sec (Mashayekhi et al. 2019a)

Because of the complexity of the requirements for producing ET excitations, optimization techniques are typically employed to produce appropriate excitations. In the optimization context, appropriate objective functions should be defined. A simple and effective objective function for simulating endurance time excitations can be defined as:

$$F(a_g) = \int_{T_{min}}^{T_{max}} \int_0^{t_{max}} \left\{ \left[ S_a(T,t) - S_{aT}(T,t) \right]^2 + \alpha \left[ S_u(T,t) - S_{uT}(T,t) \right]^2 \right\} dTdt \qquad (6)$$

where $a_g(t)$ is the acceleration time history of endurance time excitations, which is the output of minimizing this objective function, $S_{aT}(T,t)$ and $S_{uT}(T,t)$ are acceleration and displacement spectra of endurance time excitations at period $T$ and time $t$. $T_{min}$ and $T_{max}$ are the minimum and maximum considered periods. In Equation (6), $t_{max}$ is the duration of endurance time excitations for which ET record is to be simulated. $S_a(T,t)$ and $S_u(T,t)$ are, respectively, acceleration and displacement spectra of the endurance time excitation at period $T$ and time $t$. Acceleration and displacement spectra of the endurance time excitations are calculated through following equations:

$$S_a(T,t) = \max\left( |\ddot{x}(\tau) + a_g(\tau)| \right) \quad 0 \leq \tau \leq t \qquad (7)$$

$$S_u(T,t) = \max\left( |x(\tau)| \right) \quad 0 \leq \tau \leq t \qquad (8)$$

where $x(\tau)$ and $\ddot{x}(\tau)$ are displacement and acceleration time history of the single degree of freedom system with period $T$ at time $\tau$. In the objective function of Equation (6), acceleration and displacement spectra are typically considered. The constant $\alpha$ is a factor which normalizes and balances the relative weight of acceleration and displacement residuals in the objective function.

In the objective function given by Equation (6), the residuals are computed using the absolute values of differences. Alternatively, they could be also quantified with the use of relative values. Consequently, considering the dynamic characteristics of the expected ET records—the type of calculating residuals, values of $T_{max}$, $T_{min}$, and $t_{max}$, and the employed intensifying functions—diversify the definitions of ET objective function for simulating endurance time excitations. Other objective functions rather than the one mentioned in Equation (6) can also be used for simulating the endurance time excitations. For more advanced generations of ET excitations, diverse dynamic characteristics were incorporated in the relevant objective functions (Mashayekhi et al. (2018b, 2019b)).

There are various ways to define optimization variables in the simulation problem of endurance time excitations. The common approach is using acceleration data points of endurance time excitations directly as unknown parameters. The main benefit of this selection of optimization variables, which also called time domain-space, is its simplicity that comes from the fact that defining the base accelerations as the variables do not need signal decomposition.

Another more effective way for defining variable definition is to use coefficients of discrete wavelet transform (DWT) (Newland 1993). Wavelet transform decompose a signal into its frequency- and time-domain. The main differences between Fourier transform and DWT is that frequency changes in time cannot be neatly captured by Fourier analysis. Mashayekhi et al. (2018b, 2019b) investigated a new optimization space, composed of discrete wavelet coefficients, for simulating endurance time excitation functions. They showed that filtered DWT coefficients create excitations with smaller values of the objective function as well as standard deviations of simulated excitations as compared to the time-domain and DWT space. Simulating long-duration endurance time excitations, e.g. 40second endurance time excitations, requires more effort than simulating normal-duration endurance time excitations, e.g. 20 second endurance time excitations, due to the existence of a larger number of acceleration data points or optimization variables. In this case, Mashayekhi et al. (2018c) introduced another optimization approach—which is called increasing sine function—for simulation of long-duration endurance time excitations. Further researches are still required in order to find more efficient optimization spaces for simulation of such ET excitations.

There are numerous optimization algorithms for solving the optimization problem for simulation of ET excitations. However, most existing endurance time excitations are simulated by the classical optimization algorithms—for example, the trust-region-reflective method (Nozari and Estekanchi 2011) is widely used. The main drawback of such classical optimization algorithms is that they may be trapped in local minima. On the other hand, described objective functions of endurance time excitation problem seem to have many local optima due to its dynamic nature and the presence of many decision variables (typically in the order of 1000 decision variables or more). This obstacle can partly be overcome by using evolutionary algorithms, so further studies for finding appropriate evolutionary algorithms for the simulation of endurance time excitations is need. Many evolutionary algorithms have been developed to mimic natural processes for solving optimization problems such as genetic algorithm (GA), particle swarm optimization (PSO), ant colony optimization (ACO), and imperialist competitive algorithm (ICA). These evolutionary algorithms have several parameters that have to be calibrated prior to their implementation in a specific problem. Mashayekhi et al. (2019c) employed imperialist competitive algorithm (or ICA) for simulating endurance time excitations. They showed that better endurance time excitations are achieved by their proposed ICA-based evolutionary algorithm. However, the required computational time is increased by a factor of about 26 times in the proposed algorithm. This demanding computational time poses a major drawback in using evolutionary algorithms to simulate endurance time excitations when nonlinear responses are also included in the objective function (Kaveh et al. 2013). Different evolutionary optimization algorithms and the hybridization of classical and evolutionary optimization algorithms have to be examined to find better optimization frameworks for simulating endurance time excitations, which require further studies and comparisons between results obtained from the corresponding parametric studies.

Five generations of endurance time excitations have been developed so far for the ET method, where the general features of the excitations within each generation are almost identical. The distinguishing characteristics belonging to each of the five generations are as follows:

- **The first generation of ETEFs:** The theory of random vibration was employed to simulate the excitations of first generation and they were only produced to illustrate the concept of the ET method, and were not intended for use in practical applications. Consequently, these type of excitations was only used in the original work of Estekanchi et al. (2004).

- **The second generation of ETEFs:** The second generation of ET records (Nozari and Estekanchi 2011;Valamanesh et al. 2010) provided ET excitation functions for practical applications due to the acceptable accuracy of the simulated excitations. For this generation the optimization techniques were employed in simulating endurance time excitations. In addition, the linear response spectrum was included in the objective function. Classical optimization algorithms in time-domain were employed to simulate these excitations. It was later demonstrated that incorporating long periods in the objective function calculation improves the efficiency of these excitations in nonlinear response assessment despite the fact that nonlinear responses are not directly considered in the generation process. Series "ETA20a", "ETA20b", "ETA20c", "ETA20d", "ETA20e",

"ETA20f", "ETA20g" and "ETA20h" are the ET records or subcategories of the second generation of ETEFs. However, the long periods are only included to simulate some cases such as "ETA20d", "ETA20e", "ETA20f", "ETA20g" and "ETA20h" series. For a quick review on the specific characteristics of these ETEFs, readers are referred to Table 1. Endurance time excitations of these series are publicly available through the ET website (see Estekanchi (2019)).

- **The third generation of ETEFs:** In this generation, nonlinear displacement responses are included in the generation process. The series "ETA20en", "ETA20jn" and "ETA20in" belong to this generation. The letter "n" in the name of these series implies that nonlinear responses are considered in the simulation process. The series "ETA20in" and "ETA20jn" have three-component time histories—as can be seen from Table 1—and can be employed in three component analysis (Valamanesh and Estekanchi 2014). Endurance time excitations of these series are available via Estekanchi (2019).

- **The fourth generation of ETEFs:** Ground motion duration may have a significant impact on the structural responses (Hancock and Bommer 2007, Harati et al. 2019; Mashayekhi et al. (2019d, 2019e)). The motion duration consistency is directly included for simulation of the fourth generation of ET excitations. Prior to this generation, duration had not been directly incorporated in the generation process of ETEFs. In this regard, Mashayekhi et al. (2018a) included cumulative absolute velocity (CAV) in the generation process. As can be seen from Table 1, they produced series "ETA40lc" in their study. This series is also available through the ET website (see Estekanchi (2019)).

- **The fifth generation of ETEFs:** In the fifth generation of endurance time excitations, damage consistency is included and implemented in the ET generation process. Mashayekhi et al. (2018d) included hysteretic energy compatibility in the simulation process. Because damages induced on a structure are a function of both maximum displacement and absorbed hysteretic energy, incorporating hysteretic energy in the simulation process implies that damage consistency is satisfied for ET excitations of this generation. They produced series "ETA20kd" whose characteristics are concisely condensed in Table 1. Corresponding excitations are available in ET website (see Estekanchi (2019)).

Table 1. Essential specifications of the ET excitations at different generations

| ETEFs ID | No. of ET records | Generation | Target spectrum | Duration | Dynamic characteristics considered in simulation | Description |
|---|---|---|---|---|---|---|
| ETA20a | 3 | Second | Iranian National Building Code (Standard No. 2800) for soil type C | 20sec | Acceleration and displacement spectra | Periods up to 5 sec are considered in this generation |
| ETA20b ETA20c ETA20d | 3 | Second | Similar to case ETA20a | 20sec | Similar to case ETA20a | Similar to case ETA20a |
| ETA20e | 3 | Second | Average response spectra resulted from 7 records given by FEMA440 for soil type C | 20sec | Similar to case ETA20a | Long periods are considered in this generation |
| ETA20f | 3 | Second | Similar to case ETA20e | 20sec | Similar to case ETA20a | Similar to case ETA20e |
| ETA20en | 3 | Third | Similar to case ETA20e | 20sec | Nonlinear displacement | 20% improvement is achieved as compared to ETA20e |
| ETA20in-xyz | 3*3 | Third | Similar to case ETA20e | 20sec | Nonlinear displacement | This series is suitable for multi-component analysis |
| ETA20j | 3 | Third | ASCE07 design spectrum for city Tehran | 20sec | Similar to case ETA20a | This series is generated for comparing different spectrum |
| ETA40g | 3 | Third | Similar to case ETA20j | 40sec | Similar to case ETA20a | These series have long-duration ETEFs |
| ETA40lc | 3 | Fourth | Average response spectrum of 22 far-field records of FEMA P695 | 40sec | Acceleration and displacement spectra along with CAV parameter | These ETEFs are duration-consistent motions |
| ETA20kd | 3 | Fifth | Similar to case ETA40lc | 20sec | Acceleration and displacement spectra along with hysteretic energy | These ET records are suitable for damage assessment |

As can be readily understood from Table 1, several important characteristics of generated ground motions have been included in the objective functions of the available ETEFs. However, more studies are still required for simulation of more effective and reliable ET excitations and the challenge of producing more efficient and effective ETEFs is expected to remain open in coming years.

# 4. Seismic Response Assessment by Endurance Time Method

The main purpose of ET method implementation is to provide a reliable prediction of the engineering demand parameters (EDPs). In this regard, a great amount of efforts within the past decade has been made to assess the applicability of the ET method in structural seismic response prediction. This section reviews the research efforts in the seismic response assessment by the ET method. In this regard, previous studies concerning the structural type are categorized into two groups: building structures and non-building structures. The related studies are provided in following sub-sections.

## 4.1. Assessment of Building Structures

Initial studies employed a simplified structural mathematical model to assess the response of structures. As the first study, Estekanchi et al. (2007) employed the ET method in response assessment of moment and braced steel frames with linear material behavior. They compared the EDPs of ET method with results of traditional equivalent static and response spectrum seismic analysis procedures. The results were highly consistent between ET and the other employed analyses though a maximum difference value of less than 15% was observed. Moreover, Riahi et al. (2009) used the ET method in response assessment of the nonlinear single degree of freedom systems with different ductility factors, strength and damping ratios. The results of ET method matched well with the analytical outputs obtained through linear and nonlinear response analyses subjected to the real ground motions. However, major differences were seen in short period range in this case as well.

Next studies employed more realistic building structural models for the assessment of the ET method. Riahi and Estekanchi (2010) assessed the applicability of the ET method in the response estimation of steel moment frames in which the nonlinear characteristics of considered prototypes were incorporated by distributed plasticity models. In their study, the local and global demand parameters estimated by the ET method are compared with those obtained by two well-known procedures; static pushover analysis and nonlinear time history analysis. The ET method was able to estimate the results computed from real ground motions with a very high correlation value ($R^2$ coefficient of 0.96). Moreover, in another study by Estekanchi et al. (2011), the validity of ET results has been corroborated for a set of material models, including the model of elastic-perfectly plastic and models with stiffness deterioration and strength degradation. Boroujeni and Riahi (2013) demonstrated that ET analysis can properly assess the response modification factors and strength and stiffness effects in most cases related to the steel moment-resisting frames although the results of the ET method were not exactly consistent with the results obtained through the IDA framework (Vamvatsikos and Cornell 2002). To find a concise summary of the ET capability before 2014, readers are invited to study a paper by Hariri-Ardabili et al. (2014) who presented a state-of-the-art review of the ET method procedure for structural response estimation.

Recently, Mashayekhi et al. (2018d, 2019a) assessed the last generation of ET excitation functions in response estimation of building structures. In this case, considerable improvement in response prediction of the ET method was reported. They also showed that the results obtained by the ET method are quite comparable with the ones computed with the IDA procedure. Both of these studies—which are presented in this paragraph—showed about 95% accuracy in estimating the IDA results that are approximated by the ET method. As can be seen from Figure 5, it is quite recognizable that median and 16$^{th}$ fractile IDA curves can be well predicted by the ET method. This figure is extracted from an analysis conducted on an 8-story RC frame (see Bazmooneh and Estekanchi (2018) as well as Mashayekhi et al. (2019a) for more information). It is worth mentioning that the accuracy of ET method against the results found from IDA framework is also confirmed by other works that focused on structural fragility analysis (Tavazo and Ranjbaran 2017) and topics pertinent to the generation of improved ETEFs (Bai and Ou 2016; Li et al. 2019).

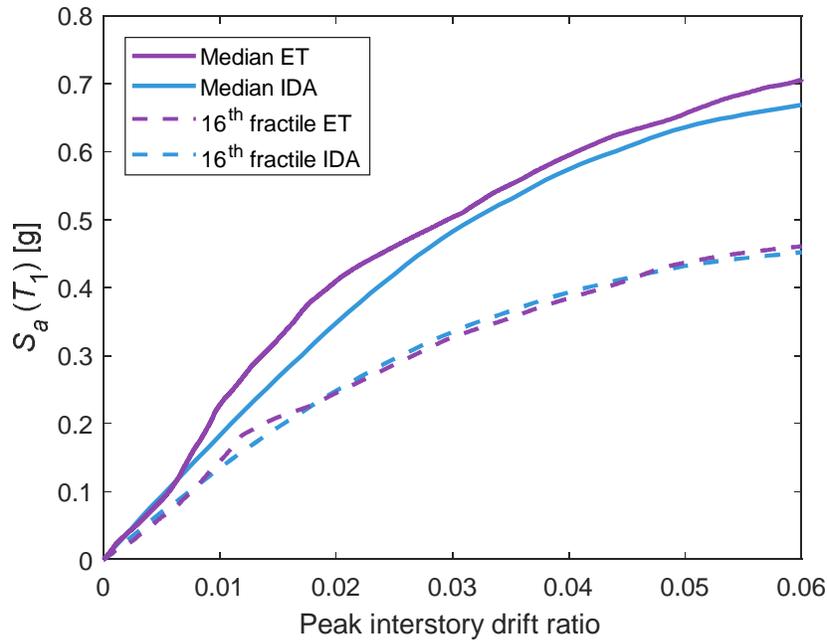

Figure 5. Response estimation using the ET method versus IDA analysis for an eight-story structure (Mashayekhi et al. 2019a)

Additional studies employed the ET method in demand prediction of special structural systems. Estekanchi et al. (2011), Vaezi et al. (2014) as well as Foyouzat and Estekanchi (2016a, 2016b) utilized the ET method in response assessment of passively controlled structures by viscous dampers, metallic dampers and friction dampers, respectively. These studies revealed a desirable performance of the ET method. It was also reported by Shirkhani et al. (2015) that ET can be reliably employed for predicting structural responses of the steel frames that are equipped with rotational friction dampers. A study focused on the optimization of the performance of type 1 fuzzy controller has also shown that the results computed with both ET and IDA methods are quite comparable (Azadvar et al. 2019). Estekanchi et al. (2018) employed the ET method in an assessment of the interaction of moment-resisting frames and shear walls in RC dual systems. In this paper, they considered evaluating the behavior of the RC dual structures according to the ET method, allowing the analysis of the behavior of the structures with different characteristics for the same seismic excitations. Moreover, Bai et al. (2018) assessed the performance of steel plate shear wall (SPSW) system by the ET method as well. They suggest a novel method for calculating ET target times which are based on the motion duration parameter.

The ET method has also been extended in the other application areas of time history analysis, which are all within the framework of the Performance-Based Earthquake Engineering (PBEE). For example, Valamanesh and Estekanchi (2013, 2014) extended the ET method in bi- and tri-directional seismic response analysis. In the latter study, a maximum difference of 15% with a 90% reduction in computational demand was observed utilizing the ET method when compared to the time history analysis that ran under seven real ground motions. Also, Mashayekhi et al. (2019a) used the ET method to estimate record-to-record variability in seismic response assessment, showing an 85-90% accuracy compared with the IDA procedure conducted using 44 real ground motions. Similarly, Bai et al. (2018, 2019) used the ET method for considering the seismic evaluation of complex systems like the ones equipped with the soil-structure interaction. In these cases, it was demonstrated that the ET method results closely follow the outputs of IDA method. To show the capability of ET method for application to more complex building models, Estekanchi et al. (2008), as well as Maleki-Amin and Estekanchi (2018), developed the ET method further to estimate the Park-Ang damage index for a number of building structures. Moreover, Rahimi and Estekanchi (2015) and Tajmir Riahi et al. (2015) established a creative and novel collapse analysis procedure by the ET method, which typically predicts the collapse capacity of the structural system with less than 15% bias as it is compared to the time history analysis. Besides, ET method is also employed in progressive collapse approaches (Nezamisavojbolaghi et al. 2017) to evaluate the contribution of infills on the amount of vertical deflection of studied frames when different column removal scenarios are considered.

### 4.2. Assessment of Non-Building Structures

The ET method was also employed in the performance assessment of non-building structures. Estekanchi and Alembagheri (2012) and Tavazo et al. (2012) used the ET method in shell structures such as liquid storage tanks. They modeled the liquid-structure interaction using the finite-element model. Compared to the outputs obtained from time history analysis, reasonable accuracy with about 0.95 coefficient of correlation has been observed in the stress prediction of the structural elements from the ET method (Estekanchi and Alembagheri 2012). Moreover, Valamanesh et al. (2011), Hariri-Ardebili and Mirzabozorg (2014), Hariri-Ardebili and Saouma (2015) as well as Furgani et al. (2019) employed the ET method for response analysis of concrete and arch dam structures. Concrete gravity dams were likewise examined by the ET method through different research works performed by Qiang et al. (2018, 2019a, 2019b). Also, Guo et al. (2017) used the ET method in multi-span highway bridge, where they modeled the pounding effect of bridge basement. In this study, the accuracy of the ET method has also been validated for bridges. Since the finite element modelling of aforementioned structures results in a rather large number of equations, the ET method will be very efficient compared to the traditional time-history analysis. For this reason, Hasani et al. (2017) and Dastan Diznab et al. (2019) utilized the ET method to reduce the demanding computational time required for the analysis of offshore structures. The latter study reported a correlation coefficient of about 0.98 between the ET method and the time history analysis. In order to gain benefits from the efficiency of the ET method in large structural systems with refined finite element models, it is reported that ET method has been used for the seismic performance evaluation of ASR-affected nuclear containment vessel (Saouma and Hariri-Ardebili 2019) and unreinforced masonry dome (Chiniforush et al. 2016) as well.

As a significant contribution, Zeinoddini et al. (2012) extended the concept of the ET method to marine engineering. They introduced the Endurance Wave Analysis (EWA) as a novel approach for nonlinear dynamic assessment of offshore structures subjected to sea waves. The wave or storm (Zeinoddini et al. 2016) analysis is similar to earthquake response analysis since the wave loads have dynamic and probabilistic nature. Also, the extreme value of such load demands is of interest to designers working in the relevant fields. In the EWA method, the offshore structure is subjected to an intensifying wave function. The wave height and its spectral density increase over time and the damage indices is monitored over time. In this regard, Diznab et al. (2014), Jahanmard et al. (2017) and Zeinoddini et al. (2018) performed additional studies and further developed the concept of EWA analysis.

## 5. Seismic Design by ET Method

The ET method is a robust platform for seismic response analysis of infrastructures. Ideally, this tool can be used not only for the seismic response assessment but also for cases relating to the structural design. In this regard, in parallel to developing the ET procedure for response assessment, various efforts have been made to develop ET-based frameworks for structural design. Previous studies concerning the design approach using the ET method are categorized into three groups for reviewing in this study, namely, performance-based, life cycle-based and value-based design. The relevant reviews are provided in the following sub-sections.

### 5.1. Performance-Based Design

Performance-based design (PBD) procedure (ASCE/SEI 41-17 (2017)) has been developed as an enhanced alternative to the conventional code-based procedures offered by ASCE-07 (2010). In the performance-based procedure, the actual performance of structures is evaluated by employing a more realistic dynamical model and a more robust response assessment tool. Moreover, using the PBD framework the seismic performance of the structures can be evaluated under multi-level of earthquake hazards. In this case, pushover procedure and time history dynamic analysis are two traditional methods for checking the design requirements. However, low accuracy of pushover procedure and high computational demand of time history analysis motivate researchers to develop alternatives to the available response analysis tools (Hariri-Ardebili et al. 2014). In this regard, the ET method provides an efficient alternative considering its acceptable accuracy and modest required computational demand. Mirzaee et al. (2010) developed a practical procedure for the application of the ET method in the performance-based design of structures. They proposed a method to find an equivalent ET target time for an arbitrary seismic hazard level. For instance, Figure 6 displays the target and existing performance curves of a hypothetical building. In this case, this building meets the Immediate Occupancy (IO) and Life Safety (LS) limit states, but it fails to satisfy the Collapse Prevention (CP) criterion. In another study, the same authors proposed a continuous mapping between ET excitation time and seismic hazard return period (Mirzaee et al. 2012). This mapping paved the way for the presentation of the ET response curve to be consistent with the median response curve computed by IDA (Vamvatsikos and Cornell 2002). In addition, based on the proposed framework, Mirzaee and Estekanchi (2015) developed a PBD-based retrofitting procedure by the ET method. Also, Radmanesh and Mohammadi (2018) decided to utilize the ET method for the PBD-based seismic evaluation of concrete special moment resisting frames designed with different seismic coefficients.

It is known that finding an economical structural design that meets the desired performance objectives needs a high number of try and error, and consequently, considerable computational time is needed in the case of conventional time history analysis. However, this issue can be overcome if the ET method is utilized as an analytical framework. In this regard, Estekanchi and Basim (2011) presented a performance-based optimum design of passively controlled structures by the ET method. They optimized the damper placement and the properties of the viscous dampers used.

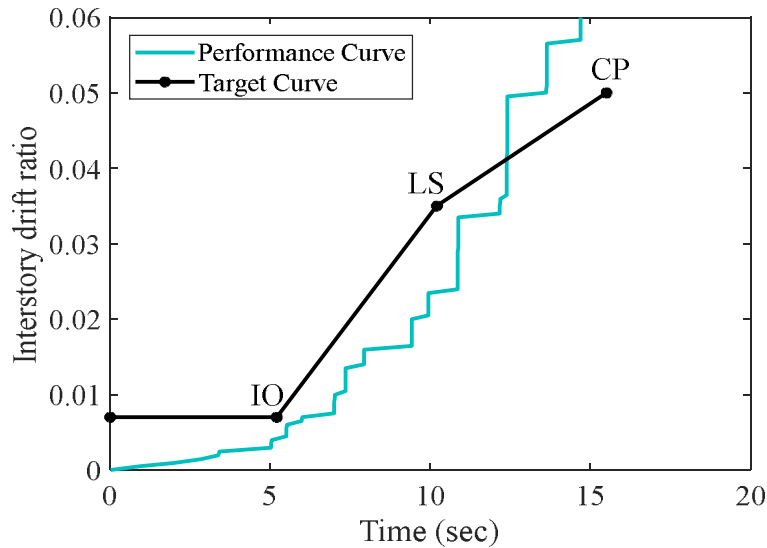

Figure 6. Target performance and existing performance curves (Mirzaee et al. 2010)

### 5.2. Life-Cycle Cost Based Design

The deficiencies in the current performance-based seismic design procedure led the researchers to develop an improved performance-based design framework (FEMA 445 (2006)). In the conventional PBD procedure, the seismic performance is expressed in a discrete and qualitative manner. The performance of non-structural components is not assessed and the level of reliability is not known (Hamburger et al. 2012). In this regard, the next-generation performance-based design has been planned to develop in ATC-58 project. In the new design framework, the performance of a building is expressed with continuous and quantitative measures. These metrics include repair cost, time of occupancy interruption, injuries and casualties. This procedure explicitly incorporates the performance of non-structural components and potential uncertainties. As a result, the next-generation performance-based design is a Life-Cycle Cost (LCC) based design (Frangopol and Soliman 2016). The LCC-based design includes the quantitative consequences as well as the construction considerations. Therefore, the structure designer should find a balance between the acceptable risk of seismic consequences and the construction costs (Matta 2017).

The design framework by the ET method has also been upgraded to incorporate the LCC as a design objective. Basim and Estekanchi (2015) developed an optimum LCC-based structural design procedure. In their proposed procedure, for each design alternative, the ET method estimates the engineering demand parameters at multi-level of seismic hazards. Then, the LCC including the construction cost and seismic consequences is evaluated based on the structural properties and results of the ET method. The trial and error procedure will be continued until the solution is converged to the best design. It is worthwhile mentioning that since the LCC-based seismic design requires the EDPs at much higher hazard levels than the PBD procedure, the efficiency of the ET method implementation would be more pronounced in this case because the ET method predicts the structural responses as a continuous function of seismic intensity by default. This framework that is based on an ET procedure is also applied to the LCC-based optimum design of lead-rubber base isolation system (Mousazadeh et al. 2019), where they reported that offered ET-based method in this study is found to be effective in diminishing the initial and life cycle cost of the considered steel frames with an acceptable accuracy and low computational effort.

### 5.3. Value Based Design

As a recently proposed design framework by the ET method, Basim and Estekanchi (2014) as well as Basim et al. (2016) introduced a Value-based Seismic Design (VBSD) concept. In their approach, the "value" parameter is considered as a more general description of design target. The value parameter incorporates the financial/economic value of seismic losses such as structural damages, damage to building contents, losses due to occupancy interruption and casualties. They employed their developed value-based framework for the design of a five-story steel moment frame by the ET method. In order to avoid encountering the very time-consuming procedure of the IDA framework in predicting the structural seismic responses required in the FEMA P-58 (2012) approach, Tafakori et al. (2017) proposed a new methodology for probabilistic seismic loss estimation that works with ET method instead of IDA analysis. They found that the suggested approach can have appropriate preciseness and efficiency when it is compared with a benchmark that is based on an IDA framework.

Mirfarhadi and Estekanchi (2020) have recently improved the framework of existing value-based design schemes. They extended the value parameter to incorporate a more comprehensive set of decision indicators as listed in Figure 7. In the new framework, first, the significant decision indicators are selected and then all of the selected indicators are evaluated and weighted using a decision-making tool. The structure is designed in a way that the maximum total value is gained, where an optimization algorithm is employed to find the optimum solution. Considering the fact that optimization procedure demands a high number of performance evaluations, employment of conventional time history analysis such as IDA surely leads to an impractically expensive design process. The ET method, however, provides a straightforward and efficient structural value-based design.

It is interesting to note that the value-based design framework, based on Figure 7, is equivalent to traditional performance-based structural design (e.g. Estekanchi and Basim 2011) if the performance-based criteria are the only selected decision indicators. Likewise, if the risk of seismic consequences and construction consideration are only taken as decision indicators, the design process will be similar to LCC-based seismic design (e.g., Basim and Estekanchi 2015). Along this line, Estekanchi et al. (2016) used only the structural resilience as the desired decision indicator and proposed a framework of resilience-based seismic design by the ET method.

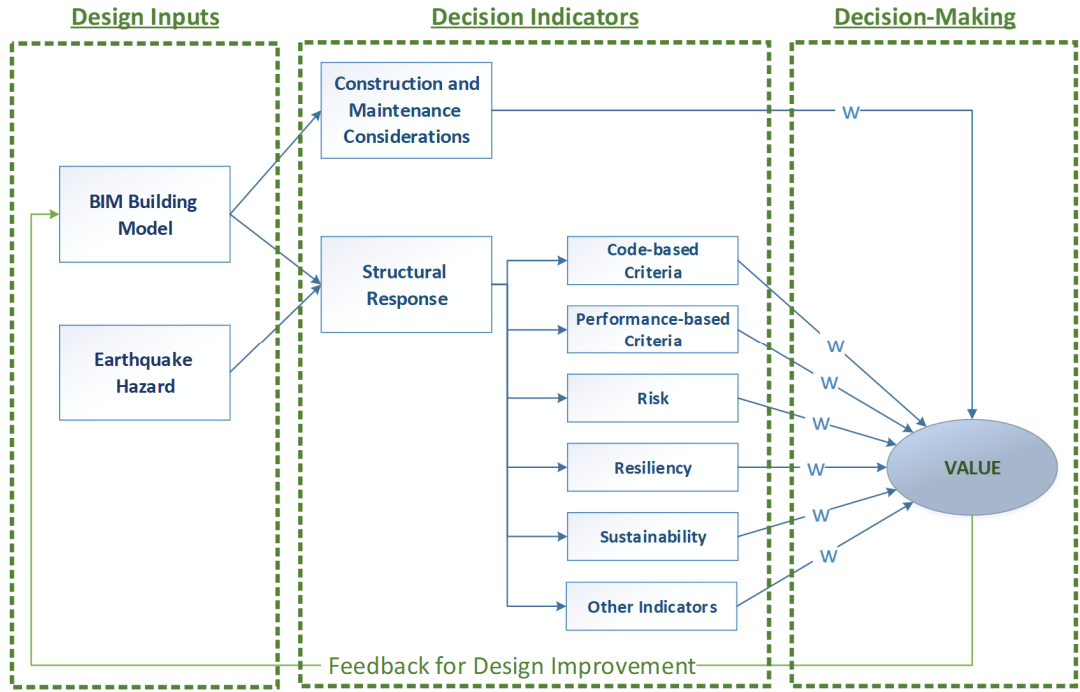
Figure 7. Extended framework of Value-based Design approach (Mirfarhadi and Estekanchi 2020)

# 6. Challenges and future developments

Past studies have demonstrated the efficiency of the ET method in comparison with conventional time history analysis in terms of reduction in analysis time, simplicity, and acceptable accuracy. However, the accuracy and efficiency of this method is highly dependent on the quality of the ETEFs that can be produced. While the ET method has been commonly utilized for seismic assessment of different structural types, using the ET method in value-based design or for the calculation of the distribution of responses is limited to a few cases since the development of the ET method in these areas is a part of an ongoing research. The ET method has the potential to be applied in different areas of earthquake engineering. Moreover, the ET method should be further improved and extended to address the following issues in near future:

I. Efficient methods for prediction of record-to-record and other sources of variability.
II. Residual displacement of buildings subjected to earthquake records.
III. Seismic response assessments of the structures that are subjected to mainshock-aftershock sequences.
IV. Effects of the near-field earthquakes (with and without near-field pulses).
V. Development of ET records adapted to be utilized as a seismic input for shaking table tests considering individual shake table limitations.

As the accuracy and efficiency of the ET method rely on the accuracy of endurance time excitations, more accurate excitations would definitely improve the reliability of this method. For all the current and future applications of the ET method, it is understandable that we may need to have more precise ETEFs which are able to satisfy our specific conditions for the problems we encounter. Some challenges in producing the endurance time excitations are as follows:

- It is essential to make use of robust optimization approaches to solve equations of new ETEFs. Producing high quality endurance time excitations is time-consuming and needs high-performance computers for solving the governing equations.

- Several optimization algorithms either classical or heuristic have been developed and are available in the literature. But many of them have not yet been applied to the problem of generating endurance time excitations. More efficient algorithms are yet to be identified and applied so that more accurate excitations can be simulated.

- The main distinguishing characteristics of simulating endurance time excitations are a large number of decision variables in their optimization problems. Investigating different optimization spaces or even creating new optimization spaces may lead to better excitations.

- Another important issue in simulating endurance time excitations is to select an appropriate objective function. Simulating objective functions may differ in form when we consider different shaking characteristics. Since objective functions have a very important impact on the resulted ETEFs, efforts should be made to propose better objective functions for the generation of new ETEFs.

- In case ET records that are intended to be used as a seismic input for shaking table tests, limitations on frequency band, maximum acceleration, velocity, and displacement as well as practical levels of accuracy should be considered.

- Duration consistency is one of the important aspects of ETEFs. This feature can be attained by incorporating appropriate intensity measures in the simulation procedure of ETEFs. These intensity measures can be then used for amplitude scaling of the expected ET excitations—both for using available ETEFs or in the production of new ones. To date, different approaches have been tried or proposed in the current literature, but working to find more efficient methods for this problem can be challenging.

## 7. Conclusions

Endurance time method is a time history dynamic analysis in which structures are exposed to predesigned intensifying acceleration time histories. This method can be used as an alternative to the conventional time history dynamic analysis. In this paper, the basic concept of the ET method was presented. The main core of the ET method is its endurance time excitations, where the reliability of the ET method results heavily relies on the quality and efficiency of these excitations. The generation process of these excitations was briefly described, and the main features of the existing endurance time excitations were presented. The challenges in the generation of endurance time excitations were also discussed. Afterwards, practical and potential applications of the ET method in areas of earthquake engineering were described briefly. In conclusion, this paper summarizes recent advances in the development of the endurance time method. Considering the simplicity and accuracy and efficiency of the Endurance Time procedure, its application is expected to become more widespread among researchers and practitioners in the field of earthquake engineering and structural engineering in near future.

Valamanesh, V., & Estekanchi, H. E. (2013). Compatibility of the endurance time method with codified seismic analysis approaches on three-dimensional analysis of steel frames. *The Structural Design of Tall and Special Buildings*, *22*(2), 144–164.

Valamanesh, V., & Estekanchi, H. E. (2014). Nonlinear seismic assessment of steel moment frames under bidirectional loading via Endurance Time method. *The Structural Design of Tall and Special Buildings*, *23*(6), 442–462.

Valamanesh, V., Estekanchi, H. E., & Vafai, A. (2010). Characteristics of second generation endurance time acceleration functions. *Scientia Iranica*, *17*(1), 53–61.

Valamanesh, V., Estekanchi, H. E., Vafai, A., & Ghaemian, M. (2011). Application of the endurance time method in seismic analysis of concrete gravity dams. *Scientia Iranica*, *18*(3), 326–337.

Vamvatsikos, D., & Cornell, C. A. (2002). Incremental Dynamic Analysis. *Earthquake Engineering & Structural Dynamics*, *31*(3), 491–514. https://doi.org/10.1002/eqe.141

Zeinoddini, M., Namin, Y. Y., Nikoo, H. M., Estekanchi, H., & Kimiaei, M. (2018). An EWA framework for the probabilistic-based structural integrity assessment of offshore platforms. *Marine Structures*, *59*, 60–79.

Zeinoddini, M., Nikoo, H. M., & Estekanchi, H. (2012). Endurance Wave Analysis (EWA) and its application for assessment of offshore structures under extreme waves. *Applied Ocean Research*, *37*, 98–110.

Zeinoddini, M., Nikoo, H. M., & Yaghubi, Y. (2016). Fragility curves of existing offshore platforms against storm loads using endurance wave analysis (EWA). *Mechanics of Structures and Materials: Advancements and Challenges - Proceedings of the 24th Australasian Conference on the Mechanics of Structures and Materials*, *1*, 903–910. https://doi.org/10.1201/9781315226460-14025